\documentclass[aps,prl,twocolumn,showpacs,groupedaddress]{revtex4}
\usepackage{graphicx}
\usepackage{color} 
\usepackage{graphicx}
\usepackage{epsfig}
\preprint{BSCCO-SW/2009}
\begin{document}
\DeclareGraphicsExtensions{.eps, .jpg}
\bibliographystyle{prsty}
\input epsf

\title{High-temperature optical spectral weight and Fermi liquid renormalization in Bi-based cuprates} 
\author{D. Nicoletti$^{1}$,  O. Limaj$^{2}$,  P. Calvani$^{1}$, G.~Rohringer$^{3}$, A.~Toschi$^{3}$, G.~Sangiovanni$^{3}$, M.~Capone$^4$, K. Held$^{3}$, S. Ono$^{5}$, Yoichi Ando$^{6}$, and S. Lupi$^{2}$}
\affiliation {$^{1}$ CNR-SPIN and Dipartimento di Fisica, Universit\`a di Roma ``La Sapienza'',
Piazzale A. Moro 2, I-00185 Roma, Italy}
\affiliation {$^{2}$ CNR-IOM, and Dipartimento di Fisica, Universit\`a di Roma ``La Sapienza'',
Piazzale A. Moro 2, I-00185 Roma, Italy}
\affiliation {$^{3}$Institute of Solid State Physics, Vienna University of Technology, 1040 Vienna, Austria}
\affiliation {$^{4}$ CNR-SMC and Dipartimento di Fisica, Universit\`a di Roma ``La Sapienza''}
\affiliation {$^{5}$Central Research Institute of Electric Power Industry, Komae, Tokyo 201-8511, Japan}
\affiliation {$^{6}$Institute of Scientific and Industrial Research, Osaka University, Ibaraki, Osaka 567-0047, Japan}

\date{\today}

\begin{abstract} 
The optical  conductivity $\sigma(\omega)$ and the spectral weight $W(T)$ of
two superconducting cuprates at optimum doping, Bi$_2$Sr$_{2-x}$La$_x$CuO$_6$
and Bi$_2$Sr$_{2}$CaCu$_2$O$_8$, have been first measured up to
500 K. Above 300 K, $W(T)$  deviates from the usual $T^2$ behavior in both compounds, even though $\sigma(\omega \to 0)$ remains larger than the Ioffe-Regel limit. The deviation is surprisingly well described  by the $T^4$ term of the Sommerfeld expansion, but its coefficients are enhanced by strong correlation, as shown by the good agreement with dynamical mean field calculations.

\end{abstract}


\maketitle

Since the discovery of high-$T_c$ cuprates, the research has been obviously focused on their "low-temperature"  properties.
However the electronic correlations, which are expected to play a central role in the low-$T$
phenomena like superconductivity and pseudogap, are also likely to affect the cuprate properties at higher $T$. In this respect, the high-$T$ behavior can provide direct information about the real nature of the fermionic excitations, which, at low-$T$, may be masked by competing ordering phenomena.
Indeed, it was  suggested that high-$T$ effects such as the violation of the Ioffe-Regel (I-R) limit for resistivity saturation and the quasi-particle (QP) thermal decoherence, are a hallmark of the same Hubbard physics which controls the low-$T$ phase diagram \cite{Gunnarsson-03}. 
Nonetheless, the high-$T$ properties of the cuprates have been scarcely investigated up to now. 
Here we present a study of the low-energy electrodynamics in two Bi-based cuprates at optimum doping from $T_c$ to 500 K, the first one of this kind to our knowledge.

We focus our investigation on the optical spectral weight
 
\begin{center}
\begin{equation}
W(\Omega,T)=\int_0^{\Omega}\sigma(\omega,T)\mathrm{d}\omega
\label{Eq1}
\end{equation}
\end{center}

\noindent 
where $\sigma(\omega,T)$ is the real part of
the $ab$-plane optical conductivity, and $\Omega$ is a cut-off frequency. $W$ is a model-independent quantity which provides important information on the evolution of the electronic dynamics with temperature \cite{Molegraaf-02,Boris,Ortolani-05,Carbone-06, Benfatto-04,Toschi-05, Norman-08}.

For $\Omega\rightarrow\infty$, the standard $f$-sum rule implies that $W$ is independent of $T$. However, useful ``restricted sum rules''  can be defined for finite $\Omega$'s. If for example $\Omega = \omega_p$ (i.e., the screened plasma frequency which in the following will be approximated with the plasma edge in the reflectivity $R(\omega)$), a tight-binding model with nearest-neighbor hopping provides:

\begin{center}
\begin{equation}
W(\omega_p,T)=-\frac{\pi e^2 a^2}{2\hbar^2V}K(T) \simeq W_0-B(\omega_p)T^2+C(\omega_p)T^4
\label{Eq2}
\end{equation}
\end{center}

\noindent
where $a$ is the lattice constant, $V$ is the sample volume, and
the second (approximate) equality in Eq. \ref{Eq2} comes from the Sommerfeld expansion of the kinetic
energy $K(T)$ up to the fourth order in $T$. In the literature, Eq.  \ref{Eq2} is typically limited to the second order ($C$ = 0), as the $T^2$-dependence of $W$ is well verified in many metals  below room temperature, including several high-$T_c$ superconductors \cite{Molegraaf-02,Ortolani-05}. Despite this ``conventional'' behavior, the cuprates show peculiar features \cite{Ortolani-05}: i) the $T^2$ dependence extends up to $\Omega$'s far larger than $\omega_p$; ii)  unlike in ordinary metals, where $W_0$ - which accounts for all the carriers in the conduction band - and the ``thermal'' coefficient $B(\omega_p)$ are governed by the same nearest-neighbor hopping rate $t_0$, in cuprates  $B \propto t_T^{-1}$ with $t_T \approx t_0/10 $. While some authors \cite{Norman-08} attribute these effects entirely to the existence of a finite cutoff, in Ref. \cite{Toschi-05} a quantitative agreement has been found between experimental data and dynamical mean field theory (DMFT) calculations, taking into account both the finite cutoff and the strong correlation present in high-$T_c$ cuprates.


\begin{figure}   \begin{center}  
\leavevmode    
\epsfxsize=8 cm \epsfbox {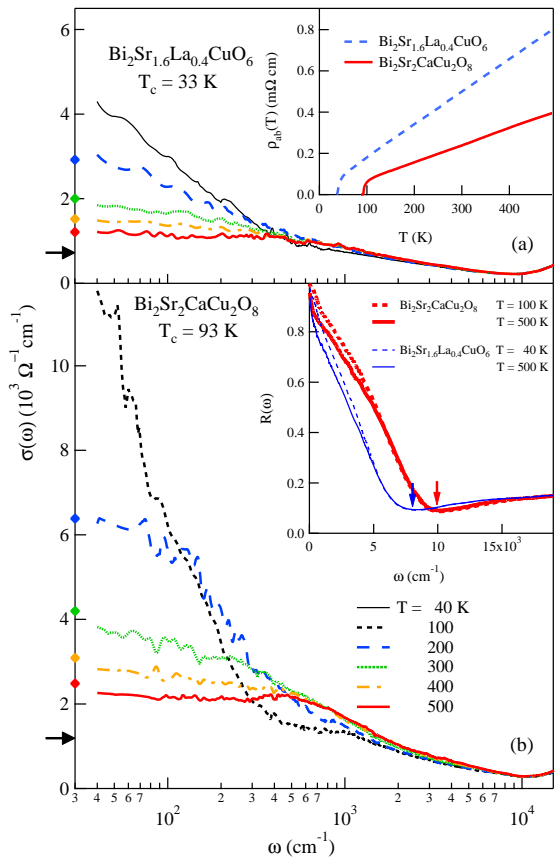}    
\caption{(Color online) (a) Normal-state $ab$-plane conductivity of 
Bi$_{2}$Sr$_{1.6}$La$_{0.4}$CuO$_{6}$ and (in the inset) resistivity of both samples. (b) Normal-state $ab$-plane conductivity of  Bi$_2$Sr$_2$CaCu$_2$O$_8$ and (in the inset) reflectivity at the two extreme temperatures of both samples, with  arrows marking their screened plasma frequencies $\omega_p$. In both main panels, the diamonds at $\omega=0$ indicate the $\sigma_{dc}$ of the sample at the same $T$ (at low $T$  both of them are out of the scale). The horizontal arrows mark the Ioffe-Regel limits to $\sigma_{dc}$ (see text).}
\label{Fig1}
\end{center}
\end{figure}


In the present paper we investigate the in-plane optical conductivity  of two cuprates in their normal phase up to 500 K, in order to study the behavior of $W(\Omega,T)$ towards the I-R limit. The samples are two single crystals at optimum doping, Bi$_2$Sr$_{1.6}$La$_{0.4}$CuO$_6$ (Bi-2201) and Bi$_2$Sr$_2$CaCu$_2$O$_8$ (Bi-2212), grown by the floating-zone (FZ) technique \cite{Ando-99}.
Bi$_2$Sr$_{2-x}$La$_x$CuO$_6$ is a single Cu-O layer cuprate, with maximum critical temperature  $T_c^{max}\simeq33$ K. In this system, whose optical spectra at low doping were reported previously \cite{Lupi-09}, optimum doping occurs at  $x=$0.4, which corresponds to $0.16$ holes per Cu \cite{Ando-00, Ono-03}.
Bi$_2$Sr$_2$CaCu$_2$O$_8$ is the well known double-layer cuprate with  $T_c^{max}\simeq93$ K.

The $ab$-plane resistivity  $\rho(T)$ is represented in the inset of Fig. \ref{Fig1}. 
Above the sharp superconducting transitions at $T_c = 33$ K for Bi-2201 and 93 K for Bi-2212, both curves  display the linear behavior typical of optimally-doped cuprates. 
The reflectivity $R(\omega)$ was measured at near-normal incidence, shortly after cleaving the samples, with a Michelson interferometer between 40 and 22000 cm$^{-1}$, at several $T > T_c$, stable within $\pm$ 1 K.
The reference in the infrared (visible) range was a gold (silver) film evaporated \textit{in situ} onto
the sample, which was mounted in a closed-cycle cryostat below room T, heated inside an optical vacuum chamber above 300 K. 
The intensity reflected both by the sample and reference was measured at every $T$, for any spectral range, in order to compensate the thermal displacements of the sample holder. The chemical stability of both samples was checked by measuring $R(\omega)$ at 300 K after every high-$T$ cycle.
$R(\omega)$ is shown at the lowest and highest temperatures in the inset of Fig. \ref{Fig1} for both Bi-2201 and Bi-2212. As shown by the vertical arrows in the inset, at any $T$ the plasma edge is at $\omega_p\simeq8000$ cm$^{-1}$ in Bi-2201 and $\omega_p\simeq10000$ cm$^{-1}$ in Bi-2212.

The real part  $\sigma(\omega)$ of the $ab$-plane optical conductivity was finally obtained from $R(\omega)$ through standard Kramers-Kronig transformations. First extrapolations of $R(\omega)$ to $\omega$ = 0  were based on Drude-Lorentz fits, which provided a $\sigma(0)$ which deviated from the  $\sigma_{dc}$  measured at the same $T$ within $\pm$ 1\%. Afterwards, these fits were adjusted exactly to $\sigma_{dc}$. The extrapolations to high frequency were based on the data of Ref. \cite{Terasaki-90} up to 40 eV and on a power law beyond this energy. The resulting  $\sigma(\omega)$
is shown in Fig. \ref{Fig1}(a) and \ref{Fig1}(b) at selected temperatures between $T_c$ and 500 K, for Bi-2201 and Bi-2212, respectively.
In both samples the edge of the lowest electronic band appears at $\omega\gtrsim10000$ cm$^{-1}$. The Drude peak in the far infrared broadens with increasing temperature, becoming a flat contribution at 500 K.


\begin{figure}   \begin{center}  
\leavevmode    
\epsfxsize=8 cm \epsfbox {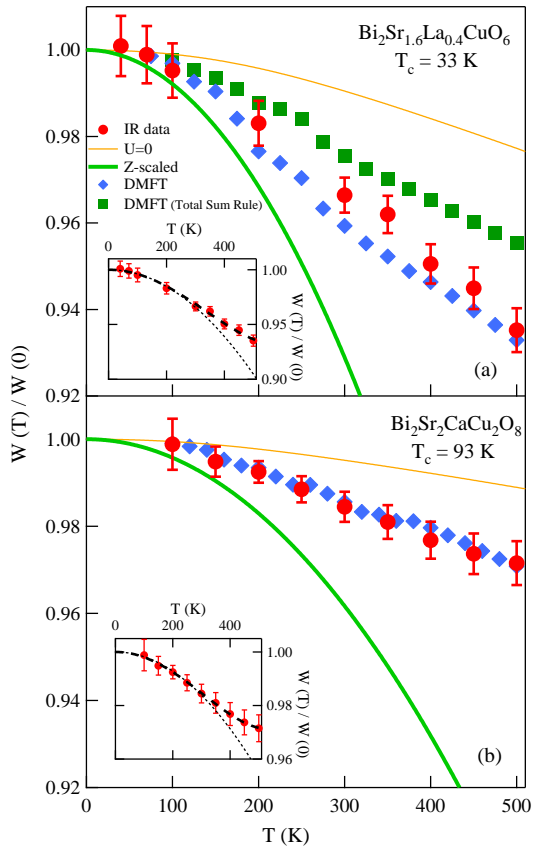}
\caption{(Color online) Temperature-dependent optical spectral weight $W(\omega_p, T)$ of optimally doped
(a) Bi-2201 and (b) Bi-2212, normalized to the (extrapolated) value at $T=0$. The IR data (red
circles) are compared with DMFT results for the restricted sum rule
(blue diamonds) of the single-band Hubbard model. Also shown are theoretical calculations for the non-interacting
system ($U=0$) and the lowest-order Sommerfeld expansion, where the
coefficient $B$ is simply rescaled by the QP DMFT weight ($Z$-scaled). 
In panel (a) DMFT results for the total sum rule are displayed for comparison (green squares).
In the inset the dotted (dashed) line indicates the fit performed on $W(\omega_p, T)$ data using Eq. \ref{Eq2} up to the second (fourth) order.}     
\label{Fig2}
\end{center}
\end{figure}


In the insets of  Fig. \ref{Fig2}, the spectral weight  $W(\omega_p)$, as obtained from $\sigma(\omega)$ by Eq. (\ref{Eq1}), is shown as a function of $T$ for both Bi-2201 and Bi-2212.
The error bars have been estimated by assuming a 1\% error on the raw $R(\omega)$ throughout
the measuring range. As one can see from the Figure, the $T^2$ dependence predicted by Eq. (\ref{Eq2}) limited to the second order (dotted line) well fits the measured  $W(\omega_p)$  only for $T\lesssim300$ K. A strong deviation
from this behavior is instead evident above room temperature for both Bi-2201 and Bi-2212.
One may wonder whether this is due to the system approaching the I-R limit, where the quasiparticle picture breaks down because the electron mean free path $\ell$ becomes comparable with the lattice constant $a$ \cite{Gunnarsson-03}. 

To evaluate the resistivity at the I-R limit, following Ref. \cite{Gunnarsson-03} one may assume a cylindrical Fermi surface of radius $k_F=\sqrt{2\pi}/a$, with $a\simeq$ 0.383 nm and height $2\pi/c$. Here, $c\simeq1.23$ nm for Bi-2201 \cite{Russo-07}  and $c\simeq0.765$ nm for Bi-2212 \cite{Kovaleva-04} is the average separation between Cu-O planes.
At the I-R limit $\ell=a$, and hence

\begin{equation}
\rho_{ab}^{I-R}=\frac{2\pi\hbar c}{e^2 k_Fa}\simeq0.055(c/a_0)\quad\mathrm{m}\Omega\cdot\mathrm{cm}
\label{IR-limit-cuprates}
\end{equation}

\noindent 
where $a_0$ is the Bohr radius.
One thus obtains $\rho_{ab}^{I-R}\simeq1.3$ m$\Omega$ cm for Bi-2201 and $\rho_{ab}^{I-R}\simeq0.8$ m$\Omega$ cm for Bi-2212.
The corresponding $\sigma_{dc}^{I-R}$ are marked in Figs. \ref{Fig1}(a) and \ref{Fig1}(b) by arrows close to the vertical axis. 
As one can see, even at the maximum temperature investigated,
$\sigma(\omega)$ is much larger than the I-R limit, indicating that a trace of coherent fermionic excitations still survives in both Bi-based systems.
  
Moreover, one finds that the deviation of $W(\omega_p, T)$ from the $T^2$ behavior is satisfactorily reproduced at all temperatures if one fits to data the
whole Eq. \ref{Eq2}, namely if one includes the $T^4$ term of the Sommerfeld expansion  (dashed lines in both insets of Fig. \ref{Fig2}). 


\begin{figure}   \begin{center}  
\leavevmode    
\epsfxsize=8 cm \epsfbox {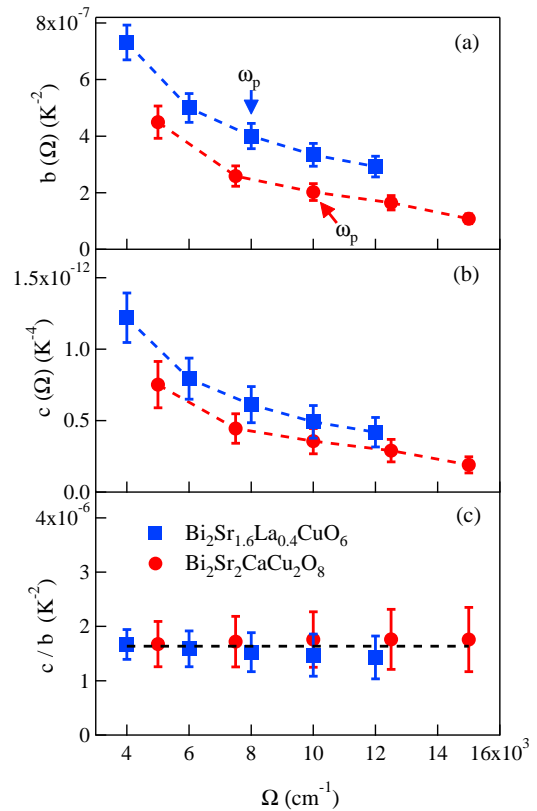}    
\caption{(Color online)
Normalized coefficients $b(\Omega)=B/W_0$, $c(\Omega)=C/W_0$,
and $c(\Omega)/b(\Omega)$ for Bi$_{2}$Sr$_{1.6}$La$_{0.4}$CuO$_{6}$ and Bi$_2$Sr$_2$CaCu$_2$O$_8$, as obtained  from the fits to $W(T, \Omega)$. The lines are guides to the eye.}
\label{Fig3}
\end{center}
\end{figure}


In order to check for the generality of such behavior,  the cutoff frequency $\Omega$
in Eq. \ref{Eq1} was varied from $\omega_p/2$ to $3\omega_p/2$, and  deviations from the $T^2$ dependence quite similar to those in the insets of Fig. \ref{Fig2} were always found. 
The resulting set of $W(\Omega,T)$ values were fit up to the $T^4$ term, and
the coefficients $B(\Omega)$ and $C(\Omega)$ were determined in terms of $W_0$. 

The results are shown in Fig. \ref{Fig3}. We obtained $b(\omega_p)=B/W_0\simeq4.0\cdot10^{-7}$ K$^{-2}$ in Bi-2201 and $b(\omega_p)\simeq2.0\cdot10^{-7}$ K$^{-2}$ in Bi-2212. 
Figure \ref{Fig3} also shows the $T^4$ coefficient $C(\Omega)$ in the Sommerfeld expansion. In both compounds, $c(\Omega) = C/W_0$ decreases with $\Omega$ like $b(\Omega)$. Therefore, within errors, the ratio $c/b$ in Fig. \ref{Fig3}(c)  is  independent of $\Omega$ and, surprisingly, also  the same in both optimally doped samples.
At the screened plasma frequency one has $c(\omega_p) = 6.1\cdot10^{-13}$ ($3.6\cdot10^{-13}$) K$^{-4}$ and
$c(\omega_p)/b(\omega_p) = (1.5 \pm 0.4)\cdot10^{-6}$ (($1.8\pm 0.5)\cdot10^{-6}$) K$^{-2}$ for Bi-2201 (Bi-2212).

The strong temperature dependence of $W$ in the low-$T$ regime and the deviations from the $T^2$ behavior at intermediate T, suggest to investigate the role of electron correlations in renormalizing the Sommerfeld coefficients $B$ and $C$. To this purpose, we have performed DMFT calculations \cite{noteED} for a single-band Hubbard model with realistic values of the hopping parameters for both Bi-based compounds \cite{notehopping}. 
Following the scheme presented in Refs. \cite{Toschi-05,Toschi-08}, such calculations qualitatively reproduce the absolute (material dependent) value of  $W$. The temperature behavior can be reproduced even quantitatively when rescaling $W$ with its (extrapolated) $T=0$ value $W_0$ \cite{Millis-08}. The comparison between experimental and DMFT values of $W(T)/W_0$ is reported in Fig. 2 for a local Coulomb interaction $U=12 t_0$ and optimal doping (0.16 holes per Cu).
The agreement with data is excellent, as DMFT results (blue diamonds in Fig. 2) capture both the strong $T^2$ dependence of $W$ at low $T$  between $0$ and $250$ K for Bi-2201 and Bi-2212 (within $\sim 3\%$  and $\sim 1.5\%$, respectively), and the deviation from the $T^2$ behavior for $T\gtrsim300$ K.
Even when the sum rule is extended up to infinity (see the DMFT calculations
for the 2201 material, green squares in Fig. \ref{Fig2})  $W$ is still
dependent on $T$. Such a dependence is certainly weaker than that obtained
for a finite cutoff, but it remains significant. This result demonstrates, that the effects of strong correlation  \cite{Toschi-05} and finite cutoff \cite{Norman-08} contribute about equally to the observed temperature dependence.
The 2201 system displays a stronger T dependence than the 2212 compound as the energy scale given both by the bare (LDA) and  renormalized (ARPES) hopping amplitude is much smaller. This does not necessitate stronger correlations since the renormalization factor (see below) is actually very similar and optical and ARPES spectroscopies agree in this respect \cite{notehopping, Toschi-08, Hashimoto-08}.
In the corresponding non-interacting system ($U=0$, thin yellow line in Fig. 2), substantially smaller values for both $B$ and $C$ are found.  This proves the major role of electronic correlations in determining the optical behavior of both Bi-based cuprates in the whole temperature range.

In a first approximation, the observed $B$ and $C$ enhancement can be related to the renormalization factor $Z$, which controls the QP bandwidth through 

\begin{equation}
1/Z= 1-\frac{\partial \Sigma(\omega=0)}{\partial \omega}
\label{Z}
\end{equation}
\noindent
Here $\Sigma$ is the DMFT self-energy and for the parameters we studied $1/Z$ ranges between $6$ and $8$. On the basis of a simple dimensional
argument one can expect that the Sommerfeld coefficients are renormalized by $Z$,
as $B\propto 1/t_0 \rightarrow 1/(Zt_0)$ and $C \propto 1/t_0^3 \rightarrow
1/(Zt_0)^3$. In Fig. 2,  both DMFT and experimental results are compared with curves
obtained by this simple rescaling of the non-interacting $B$ coefficient by
$Z$. 
The small deviations observed up to 250 - 300K (i.e. for the coefficient $B$ \cite{noteB}) can be ascribed to the $T$-dependence of the chemical potential and to the smearing of the Van Hove singularity due to correlation \cite{Toschi-08}.
On the other hand, analytical calculations show that the simple dimensional renormalization cannot be applied to the coefficient $C$, since its value \cite{noteC} also depends on frequency- and temperature-dependent scattering terms of $\Sigma$, which we found to be substantial in our DMFT calculation.

In conclusion, we have 
measured for the first time the behavior of the restricted optical sum rules up to 500 K for two optimally-doped Bi-based cuprates. The usual $T^2$ behavior does not hold above room $T$ and, for any cutoff choice, 
the experimental results can be described in terms of an effective Sommerfeld expansion up to the $T^4$ term. The large values of the expansion coefficients controlling the $T^2$ and $T^4$ terms imply (i) a strong temperature dependence in the relevant temperature range and (ii) a rather low $T$ at which deviations from the quadratic behavior become appreciable.
The validity of a Sommerfeld expansion (though for strongly correlated electrons) from $T_c$ to 500 K
is a challenging result as it indicates that the low-energy fermionic excitations in both optimally doped cuprates can be described in terms of a renormalized Fermi liquid. This result calls for further work aimed at understanding its relation with the well known deviations from the Fermi-liquid behavior attributed to the presence of a Quantum Critical Point in the phase diagram of cuprates \cite{QCP}.

M.C. activity is funded by the European Research Council under FP7/ERC Starting Independent Research Grant "SUPERBAD" (Grant Agreement n. 240524) and by MIUR PRIN 2007 Prot. 2007FW3MJX003. Work in Vienna was supported by the FP7 EU network MONAMI.



\begin{thebibliography}{}
\bibitem{noteED} Exact Diagonalization with $6-7$ sites has been used as impurity solver for DMFT calculations.
\bibitem{notehopping} The hopping parameters adopted \cite{Hashimoto-08, Pavarini-01, Toschi-08} are $t_0=$ 250 (350) meV  (nearest-neighbor), $t_1/t_0=$ -0.2 (-0.3) (diagonal), $t_2/t_1=$ -0.5 (0.0) (second nearest neighbor) for  Bi-2201 (Bi-2212).
\end{thebibliography}
\end{document}